\documentstyle [12pt]{article}
\def\fnote#1#2{\begingroup\def\thefootnote{#1}\footnote{#2}\addtocounter
{footnote}{-1}\endgroup}
\def\inbar{\vrule height1.5ex width.4pt depth0pt}
\def\IB{\relax{\rm I\kern-.18em B}}
\def\IC{\relax\,\hbox{$\inbar\kern-.3em{\rm C}$}}
\def\ID{\relax{\rm I\kern-.18em D}}
\def\IE{\relax{\rm I\kern-.18em E}}
\def\IF{\relax{\rm I\kern-.18em F}}
\def\IG{\relax\,\hbox{$\inbar\kern-.3em{\rm G}$}}
\def\IH{\relax{\rm I\kern-.18em H}}
\def\II{\relax{\rm I\kern-.18em I}}
\def\IK{\relax{\rm I\kern-.18em K}}
\def\IL{\relax{\rm I\kern-.18em L}}
\def\IM{\relax{\rm I\kern-.18em M}}
\def\IN{\relax{\rm I\kern-.18em N}}
\def\IO{\relax\,\hbox{$\inbar\kern-.3em{\rm O}$}}
\def\IP{\relax{\rm I\kern-.18em P}}
\def\IQ{\relax\,\hbox{$\inbar\kern-.3em{\rm Q}$}}
\def\IR{\relax{\rm I\kern-.18em R}}
\def\IT{\relax{\rm I\kern-.18em T}}
\def\ZZ{\relax{\sf Z\kern-.4em Z}}

   \def\cM{{\cal M}}
\def\cN{{\cal N}}

\def\fnote#1#2{\begingroup\def\thefootnote{#1}\footnote{#2}\addtocounter
{footnote}{-1}\endgroup}
\def\beq{\begin{equation}}
\def\eeq{\end{equation}}
\def\bea{\begin{eqnarray}}
\def\eea{\end{eqnarray}}
\def\lleq#1{\label{#1}\eeq}

\def\tabroom{\hbox to0pt{\phantom{\Huge A}\hss}}
\def\notin{\ \hbox{{$\in$}\kern-.51em\hbox{/}}}

\begin{document}

\hfill {
{NSF--ITP--93--30}}

\hfill {
{HD--THEP--93--05}}
\vskip .9truein
\centerline{\large \bf Critical Strings from Noncritical Dimensions:}
\centerline{\large \bf A Framework for Mirrors of Rigid Vacua
            \fnote{\diamond}{Based in part on a talk presented at the
                  Texas/PASCOS meeting, Berkeley, CA, 1992.}
            \fnote{*}{This work is supported in part by
                        NSF Grant PHY--89--04035}}

\vskip .5truein

\centerline{\sc Rolf Schimmrigk
            \fnote{\dagger}{Email address: schimmrigk@sbitp.bitnet,
                             q25@vm.urz.uni-heidelberg.bitnet}}
\vskip .2truein
\centerline{\it Institute for Theoretical Physics, University of California}
\centerline{\it Santa Barbara, CA 93106, USA}
\centerline{and}
\centerline{\it Institut f\"ur Theoretische Physik, Universit\"at Heidelberg}
\centerline{\it Philosophenweg 16, 6900 Heidelberg, FRG}

\vskip 1truein
\centerline{\bf ABSTRACT}

\vskip .2truein
\noindent
The r\^ole in string theory of manifolds of complex dimension
$D_{crit}+2(Q-1)$ and
positive first Chern class is described. In order to be useful for
string theory the first Chern class of these spaces has to satisfy
a certain relation. Because of this condition the cohomology groups of
such manifolds show a specific structure. A group that is particularly
important is described by $(D_{crit} + Q-1, Q-1)$--forms because
it is this group which contains the higher dimensional counterpart
of the holomorphic $(D_{crit}, 0)$--form that figures so prominently
in Calabi--Yau manifolds.
It is shown that the higher dimensional manifolds do
not, in general, have a unique counterpart of this holomorphic
form of rank $D_{crit}$. It is also shown that these manifolds lead,
in general, to a number of additional modes beyond the
standard Calabi--Yau spectrum. This suggests that not only
the dilaton but also the other massless string modes, such as the
antisymmetric torsion field, might be relevant for
a possible stringy interpretation.

\renewcommand\thepage{}
\vfill  \eject

\baselineskip=15.3pt  
\parskip .05truein
\parindent=20pt
\pagenumbering{arabic}

\vskip .5truein

\noindent
\centerline{\bf INTRODUCTION}

\noindent
There are many reasons why the present understanding of
string compactification is unsatisfactory. One that has  recently
attracted attention is the problem of finding the proper
framework of mirror symmetry.

According to standard lore it is believed that in left--right symmetric
compactifications without torsion
the internal space is described by a compact manifold
which is complex, K\"ahler, and admits a covariantly constant spinor,
 i.e. it has vanishing first Chern class, so--called Calabi--Yau manifolds.
Such theories lead to a simple spectrum which is described,
in part, by the cohomology of the manifold, parametrized by
its two independent Hodge numbers
$h^{(1,1)}$ and
$h^{(2,1)}$ which parametrize the number of antigenerations and
generations, respectively, that are observed in low energy physics.

It is also believed that this class of string vacua features a symmetry,
mirror symmetry, which has been discovered in the context of
Landau--Ginzburg vacua$^1$
\fnote{\P}{ It has been shown that this symmetry is not accidental:
by a combination of orbifolding and fractional transformations$^{2}$
a mirror construction can be established between a priori independent
pairs of Landau--Ginzburg
theories with opposite spectrum.} and, independently$^3$
 in the context of a subclass
of Landau--Ginzburg theories corresponding to orbifolds of
exactly solvable minimal tensor models.
The effect of this symmetry is that for each string vacuum with
$h^{(1,1)}$ antigenerations and $h^{(2,1)}$ generations there
exists a mirror vacuum for which these numbers are exchanged.
This operation thus flips the Hodge diamond
along the offdiagonal.

This symmetry creates a puzzle.
There are well--known Calabi--Yau vacua which are rigid, i.e. they do
not have string modes corresponding to complex deformations of the
manifold.
 Since mirror symmetry exchanges complex deformations and K\"ahler
deformations of a manifold
it would seem that the mirror of a rigid Calabi--Yau manifold cannot be
K\"ahler and hence does not describe a consistent string vacuum.

Here I review a recent analysis$^4$ that throws light on this
problem and present further results.
 The construction involves manifolds which are of complex
dimension $(D_{crit}+2k)$ with a positive first Chern class, which
is quantized in
multiples of the degree of the manifold. Thus they do not describe,
a priori, consistent string groundstates. Surprisingly, however, it is
possible to derive from these higher dimensional
manifolds the spectrum of critical string vacua, independently of
whether or not a Calabi--Yau
manifold can be associated to these vacua. This
can be done not only for the generations but also for the antigenerations.
For particular types of
these new manifolds it is furthermore possible to construct the
corresponding
$D_{crit}$--dimensional Calabi--Yau manifold directly from the
$(D_{crit}+2k)$--dimensional space.

This new  class of manifolds is, however, not in one to one correspondence
with the class of Calabi--Yau manifolds because it contains manifolds which
describe string vacua
that do not contain massless modes corresponding to antigenerations.
It is precisely this new type of manifold that is
needed in order to construct mirrors of rigid Calabi--Yau manifolds
 without generations.

It turns out that the spectrum of the critical vacuum $V_{crit}$,
parametrized by (generalized) cohomology groups
 is embedded in the Hodge diamond
 of the noncritical manifold $M_{D_{crit}+2(Q-1)}$
\beq
H^{(p,q)}(V_{crit}) \subset
H^{(p+(Q-1),q+(Q-1))}(M_{D_{crit}+2(Q-1)}).
\eeq
Since a CY manifold is defined by the existence of a holomorphic
$(D_{crit},0)$--form one might have expected the noncritical spaces
to be characterized by the existence of a unique
 ${\small (D_{crit}+(Q-1),(Q-1))}$--form.
 This is not true. In general the  Hodge--diamond for manifolds
of odd complex dimension $(2p+1)$ leads to
\beq
b_{2p+1}=2n+2h^{(p+1,p)}
\eeq
where $n$ is some integer larger than 0 that may exceed 1. The reason
why $n$ may, and should allowed to, be larger than 1 will become
clear below.

Finally, it will be shown that the diagonal cohomology groups
$H^{(p,p)}$ of the higher dimensional manifolds have, in general
dimensions larger than 1 and lead to a number of
modes beyond those accounted
for by the generations and antigenerations of the critical vacua.
Hence it appears that a possible stringy interpretation will not
only involve the dilaton but also the other massless string modes,
such as the antisymmetric tensor field, describing torsion.

\vskip .3truein
\noindent
\centerline{\bf HIGHER DIMENSIONAL MANIFOLDS WITH POSITIVE }
\centerline{\bf QUANTIZED FIRST CHERN CLASS}
\vglue 0.2cm
\noindent
Consider the class of manifolds of complex dimension $N$ embedded in a
weighted projective space $\IP_{(k_1,\dots, k_{N+2})}$ as hypersurfaces
defined by the zero locus of a transverse polynomial
$p(z_1,\dots,z_{N+2})$ of degree $d$.
Here the $k_i$ are the weights of the ambient weighted projective space.
Assume that for the hypersurfaces the weights $k_i$ and
the degree $d$ are related via the constraint
\beq
\sum_{i=1}^{N+2} k_i =  Qd,
\lleq{pquant}
where $Q$ is a positive integer
\fnote{\ddag}{This definition is rather
          natural in the context of the theory of Landau--Ginzburg
          string vacua with an arbitrary number of scaling fields. A
         particular simple manifold in this class, the cubic 7--fold
       $\IP_8[3]$, has been the subject of recent investigations
        of Candelas, Derrick and Parkes$^{5,6}$ and Vafa$^7$.}.
Because of (\ref{pquant}) the first
Chern class of such hypersurfaces is given by
\beq
c_1(M_{N,d}) =(Q-1)~c_1(\cN)
\lleq{c1quant}
where $c_1(\cN)=dh$ is the first Chern class of the normal bundle
$\cN$ of
the hypersurface $M_{N,d}$ and $h$ is the pullback of the K\"ahler form
${\rm H}\in {\rm  H}^{(1,1)}\left(\IP_{(k_1,\dots,k_{N+2})}\right)$
of the
ambient space. Therefore the first Chern class is quantized in multiples
of the degree of the hypersurface $M_{N,d}$.

It turns out that these spaces are closely
related to string vacua of complex critical dimension
\beq
D_{crit} = N-2(Q-1)
\eeq
 and that for certain subclasses
of hypersurfaces of type (\ref{pquant}) it is possible to construct
Calabi--Yau manifolds $M_{CY}$ of dimension $D_{crit}$  and complex
codimension $codim_{\IC} (M_{CY}) = Q$
directly from these manifolds. In terms of the critical dimension and
the codimension the class of manifolds can be
described as projective configurations
\beq
\IP_{(k_1,\dots ,k_{(D_{crit}+2Q)})}
\left[\frac{1}{Q}\sum_{i=1}^{D_{crit}+2Q} k_i\right] .
\lleq{newmfs}

As mentioned already in the introduction the class of spaces defined by
(\ref{newmfs}) contains manifolds with no antigenerations
and hence it is necessary to have some way other than Calabi--Yau
manifolds to represent string groundstates if one wants to relate
them to higher dimensional manifolds. One possible way
to do this is to associate them to
Landau--Ginzburg theories which are characterized by their chiral
ring structure encoded in the superpotential.

In certain benign situations the subring of monomials
of charge 1 in the chiral ring describes the generations of the
vacuum$^8$. For this to hold at all it is important that
the GSO projection
is the canonical one with respect to the cyclic group $\ZZ_d$,
the  order of which is the degree $d$ of the superpotential.
Thus the generations are easily derived for this subclass of
 theories in (\ref{newmfs})  because the polynomial ring is identical
to the chiral ring of the corresponding Landau--Ginzburg theory.
In general the singularities of these spaces have to be resolved and
the resolutions contribute to the complex deformations as will become
clear below.

It remains to consider the second cohomology.
For manifolds with positive first Chern class and therefore for the
manifolds under discussion all of the second cohomology resides in
$H^{(1,1)}$.
The simplest example in the class (\ref{newmfs}), $\IP_8[3]$, already
shows however that there is a mismatch between the K\"ahler sector
of the higher dimensional manifolds and that of the critical manifolds:
for $\IP_8[3]$ $h^{(p,p)}=1$, $0\leq p \leq 7$ whereas the critical
vacuum, described by the tensor model $1^9$, has no K\"ahler
deformations at all.
Since the theory does not contain modes
corresponding to (1,1)--forms it appears that a potential manifold
cannot be K\"ahler and hence not projective. Thus it seems that the
7--dimensional manifold $\IP_8[3]$ whose polynomial ring is
identical to the chiral ring
of the LG theory is merely useful as an auxiliary device
in order to describe one sector of the critical LG string vacuum:
even though there exists a precise identity between the Hodge numbers
in the middle cohomology group of the higher dimensional manifold
and the middle
dimension of the cohomology of the Calabi--Yau manifold this is not
the case for the second cohomology group.

\vskip .3truein
\noindent
\centerline{\bf NONCRITICAL MANIFOLDS AND CRITICAL VACUA}

\noindent
Even though for $N>3$ the manifolds (\ref{newmfs})
 clearly are not  of Calabi--Yau type they do in
fact encode important information about string vacua.
The evidence for this is twofold. First it is possible to derive
from these higher dimensional manifolds the massless spectrum
of critical vacua. Furthermore it can be shown that for certain
subclasses
of hypersurfaces of type (\ref{newmfs}) it is possible to construct
Calabi--Yau manifolds of dimension $D_{crit}$  and complex
codimension $codim_{\IC} (M_{CY}) =Q$
directly from these manifolds.

The basic idea$^{4}$ is that the critical string physics has its origin
in the singularity structure of the noncritical manifolds
(\ref{newmfs}). In particular the antigenerations are generated by
the singularities.

This resonates with the observation that the noncritical 7--fold
$\IP_8[3]$, associated to the $1^9$ model, is not only quasismooth but
smooth and hence no antigenerations are expected. The same is true
for the other `non--K\"ahler' type spectra that are known.
Vacua without antigenerations are rather exceptional however; the vast
majority groundstates will have both sectors, generations and
antigenerations.
The idea just described to derive the antigenerations works for
higher dimensional manifolds corresponding to rather different types
of critical vacua. I have discussed previously several
simple classes in some detail$^{4,9}$.
The simplest picture that emerges from those constructions is that
of a fibered submanifold embedded in the higher dimensional manifold
where the base and the fibres are determined by the
singular sets of the ambient manifold. The Calabi--Yau manifold itself
is a hypersurface embedded in this fibered  submanifold.

In more complicated manifolds the singularity structure will consist
of hypersurfaces whose fibers and/or base are fibered in turn,
leading to
an iterative construction. The fibered
submanifold to be considered will, in
those
cases, be of codimension larger than one and the Calabi--Yau
manifold will be described by a submanifold with codimension
larger than one as well. To illustrate this point consider the
9--fold
\beq
\IP_{(5,5,6,6,6,4,4,4,8,8,8)}[16] \ni
\{ \sum_{i=1}^2 \left(u_i^2v_i+v_i^2w_i+w_i^2x_i +x_i^2\right)
   + v_3^2w_3+w_3^2x_3 +x_3^2 =0\}.
\eeq
The $\ZZ_2$--fibering leads to the split
$\IP_1 \times \IP_{(3,3,3,2,2,2,4,4,4)}$
which in turn leads to a further $\ZZ_2$ split
$\IP_1 \times \IP_2 \times \IP_{(1,1,1,2,2,2)}$
which finally leads to
\beq
\matrix{\IP_1 \cr \IP_2\cr \IP_2 \cr \IP_2\cr}
\left[\matrix{2 &0 &0 &0\cr
              1 &2 &0 &0\cr
              0 &1 &2 &0\cr
              0 &0 &1 &2\cr}\right] \ni
\left\{ \begin{array}{c l}
p_1 =& \sum u_i^2v_i =0 \\
p_2 =& \sum v_i^2w_i =0 \\
p_3 =& \sum w_i^2x_i =0 \\
p_4 =& \sum x_i^2 =0.
         \end{array}
\right\}
\eeq
Thus the 9--fold fibers iteratively and the splits of the polynomial
$p$ are dictated by the fibering.

The construction exemplified above was shown$^{4}$ to apply to an
infinite series of noncritical manifolds of arbitrary critical
dimension. A further infinite class of interest are the spaces
\bea
\begin{tabular}{l l}
$\IP_{(1,1,....,1,\frac{n+1}{2},\frac{n+1}{2})}[n+1]$,
&$n+1$ even\tabroom \\
$\IP_{(2,2,....,2,n+1,n+1)}[2(n+1)]$, &$n+1$ odd.
\end{tabular}
\eea
of dimension $(n+1)$.
For $(n+1)$ odd the $\ZZ_2$ singular set is a
Calabi--Yau manifold and the $\ZZ_{n+1}$ singular set consists of
two points. Hence the higher dimensional space leads to two
copies of the CY manifold
\beq
\IP_n[n+1],~~~~ n \in \IN
\eeq
embedded in ordinary projective space.

The simplest case is $n=2$ for which the resolution of the
orbifold singularities of the noncritical 3--fold
\beq
\IP_{(2,2,2,3,3)}[6]
\eeq
leads to two independent Hodge numbers $h^{(1,1)}=4$, $h^{(2,1)}=2$
and hence the Hodge diamond
      contains {\it twice} the Hodge diamond of the torus, as it
must, according to the geometrical picture described above.
Similarly $\IP_{(2,2,2,2,2,5,5)}[10]$ leads
to two copies of the critical quintic.

The first important point here is the fact that $h^{(D_{crit}+Q-1,Q-1)}=2$
in this case and hence this is an example of the phenomenon mentioned
in the introduction that  $h^{(D_{crit}+Q-1,Q-1)}$ is not always unity.

Furthermore, because $h^{(1,1)}=4$,  this example shows that, in
general, there exists more than one `remaining' mode, not accounted for
by the generations and antigenerations of the critical vacuum. This
is different from the structure of smooth example $\IP_8[3]$ which
contains only one additional field. In general, then, we have to
expect that not only the dilaton but also other string modes, such
as torsion, will play a role in a possible stringy interpretation.

A general relation between the cohomology of the critical vacuum
(NOT described by a Calabi--Yau manifold in general) and the
cohomology of the higher dimensional space emerges:
\beq
H^{(p,q)}(V_{crit}) \subset H^{(p+Q-1,q+Q-1)}(\cM_{D_{crit}+2(Q-1)})
\eeq
 and the constructions reviewed here$^{4,9}$ provide a
geometrical framework for this projection indepently of whether or not
it is possible to associate a Calabi--Yau manifold to the critical
vacuum $V_{crit}$.

\vskip .3truein
\noindent
\centerline{\bf CONCLUSIONS}

\noindent
We have seen that the class of higher dimensional K\"ahler
manifolds of type (\ref{newmfs}) with positive first Chern class,
quantized in a particular way,
generalizes the framework of Calabi--Yau vacua in such a way
as to be able to account for   mirror symmetry: For
particular types of noncritical manifolds Calabi--Yau manifolds of
critical dimension are embedded algebraically in a fibered submanifold.
For string vacua which cannot be described by critical
K\"ahler manifolds and which are mirror candidates of rigid Calabi--Yau
manifolds the higher dimensional manifolds still lead to the
spectrum of the critical vacuum and a rationale emerges which explains
why a Calabi--Yau representation is not possible in such theories.
Thus these manifolds of dimension $c/3 +2(Q-1)$
 define an appropriate framework in which to discuss mirror symmetry.

There are a number of important consequences that follow from the
previous results. First it should be realized that the relevance
of noncritical manifolds necessitates the generalization of a
conjecture regarding the relation between (2,2) superconformal field
theories of central charge
$c=3D$, $D\in \IN$, with N$=$1 spacetime supersymmetry on the one hand
and K\"ahler manifolds
of complex dimension $D$ with vanishing first Chern class on the other.
It was suggested by Gepner that this relation is 1--1.
It follows from the results above$^{4,9}$ that
instead superconformal theories of the above type are in correspondence
with K\"ahler manifolds of dimension $c/3 +2(Q-1)$ with a first Chern
class quantized in multiples of the degree.

A second consequence is that for a large class of Landau--Ginzburg
theories a new canonical prescription emerges for the construction
of the critical manifold, if it exists, directly from the 2D
field theory.

Finally, it should be mentioned that the methods of toric geometry,
used in a  recent mirror construction by Batyrev$^{10}$, are not
restricted to Calabi--Yau manifolds. Hence even though Batyrev's
construction applies only to manifolds defined by one polynomial
in a weighted projective space, or products thereof, the
constructions$^4$ reviewed here may lead to the possibility of
extending Batyrev's results to Calabi--Yau manifolds of codimension larger
than one by proceeding via noncritical manifolds.

\vskip .3truein
\noindent
\centerline{\bf ACKNOWLEDGEMENT}

\noindent
I thank Vitja Batyrev, Per Berglund, Philip Candelas, Ed Derrick
and S.S. Roan for
discussions.

\vskip .3truein
\noindent
\centerline{\bf REFERENCES}
\noindent
\begin{enumerate}
\item CANDELAS, P, M.LYNKER \& R.SCHIMMRIGK. 1990.
 Nucl.Phys. {\bf B341}:383
 \item LYNKER,M \& R.SCHIMMRIGK. 1990. Phys.Lett. {\bf 249B}:237
 \item GREENE, B.R. \& R.PLESSER. 1990. Nucl.Phys. {\bf B338}:15
 \item SCHIMMRIGK, R., {\it Critical String Vacua from Noncritical
 Manifolds: A Novel Framework for String Compactification},
 HD--THEP--92--29 preprint
\item CANDELAS, P., Talk at the Workshop on Geometry and Quantum
 Field Theory, Baltimore, March 1992
 \item CANDELAS, P., E.DERRICK \& L.PARKES., {\it The Mirror of a
 Rigid Manifold}, UTTG--24--92 preprint
  \item VAFA, C., {\it Topological Mirrors and Quantum Rings},
 HUTP--91/A059 preprint
 \item CANDELAS, P., 1988. Nucl.Phys. {\bf 298}:458
 \item SCHIMMRIGK, R., {\it Noncritical Dimensions for Critical String
 Theory: Beyond the Calabi--Yau Frontier}, HD--THEP--92--47 preprint
 \item BATYREV, V.V., {\it Dual Polyhedra and Mirror Symmetry for
 Calabi--Yau Hypersurfaces in Toric Varieties}, University of Essen
 preprint; {\it Variations of the Mixed Hodge Structure of Affine
 Hypersurfaces in Algebraic Tori}, University of Essen preprint
\end{enumerate}

\end{document}